\documentclass[
jcp,
reprint%
]{revtex4-1}
\usepackage{graphicx}
\usepackage{dcolumn}
\usepackage{bm}
\usepackage{color} 
\usepackage{amsmath,amssymb}
\usepackage [latin1]{inputenc}
\usepackage{epsfig}
\usepackage{bbold}
\usepackage{url}
\usepackage{subfig}
\usepackage{lineno}
\usepackage{ulem}
\usepackage{gensymb}
\usepackage{caption}

\usepackage{graphicx}

\usepackage[ruled,vlined]{algorithm2e}

\def\ket#1{\left|#1\right\rangle}

\newcommand{\norm}[1]{\left\lVert#1\right\rVert}

\newcommand*{\citen}[1]{%
  \begingroup
    \romannumeral-`\x 
    \setcitestyle{numbers}%
    \cite{#1}%
  \endgroup   
}

\usepackage{listings}

\usepackage[colorlinks=true,citecolor={blue},urlcolor={red},linkcolor={blue}]{hyperref}

\begin{document}
\title{Optimized Quantum Phase Estimation for Simulating  
Electronic States  in Various Energy Regimes.}

\author{Christopher Kang}
\affiliation{University of Washington, Seattle, WA, 98195, United States of America}
\author{Nicholas P. Bauman}
\affiliation{Physical and Computational Sciences Directorate, Pacific Northwest National Laboratory, Richland, Washington 99354, United States of America}
\author{Sriram Krishnamoorthy}
\affiliation{Physical and Computational Sciences Directorate, Pacific Northwest National Laboratory, Richland, Washington 99354, United States of America}
\author{Karol Kowalski}
\email{karol.kowalski@pnnl.gov}
\affiliation{Physical and Computational Sciences Directorate, Pacific Northwest National Laboratory, Richland, Washington 99354, United States of America}

\date{\today}

\begin{abstract}
    While quantum algorithms for simulation exhibit better asymptotic scaling than their classical counterparts, they currently cannot be implemented on real-world devices. Instead, chemists and computer scientists rely on costly classical simulations of these quantum algorithms. In particular, the quantum phase estimation (QPE) algorithm is among several approaches that have attracted much attention in recent years for its genuine quantum character. However, it is memory-intensive to simulate and intractable for moderate system sizes. This paper discusses the performance and applicability of \texttt{QPESIM}, a new simulation of the QPE algorithm designed to take advantage of modest computational resources. In particular, we demonstrate the versatility of \texttt{QPESIM} in simulating various electronic states by examining the ground and core-level states of H$_2$O. For these states, we also discuss the effect of the active-space size on the quality of the calculated energies. For the high-energy core-level states, we demonstrate that new QPE simulations for active spaces defined by 15 active orbitals significantly reduce the errors in core-level excitation energies compared to earlier QPE simulations using smaller active spaces. 
\end{abstract}

\maketitle

\section{Introduction}

The emergence of quantum computing technologies promises to provide new modeling tools that eliminate known deficiencies of many-body formulations designed to take advantage of classical computing. Among these, one can highlight the exponential growth in the computational demands for methods that provide the so-called full configuration interaction (FCI)  level of accuracy, which is needed to provide the quantitatively correct level of accuracies required in predictive computational chemistry. Another well-known deficiency is an inherent bias (biases), or limited applicability, of all approximations used in quantum chemistry. For example, in the area of wave function based correlated methods, the applicability of single-reference many-body perturbation theory (MBPT) or single-reference coupled-cluster (CC) methods
\cite{coester58_421,coester60_477,cizek66_4256,paldus72_50,purvis82_1910,jorgensen90_3333,paldus07,crawford2000introduction,bartlett_rmp}
is limited by the quality of the so-called reference function, i.e., a  Slater determinant that provides a zeroth-order approximation to  the exact ground-state wave function. 
A typical way of manifesting these problems is associated with the nonuniform level of accuracy of calculated energies on the ground-state potential energy surfaces. For example, while MBPT methods provide a reasonable level of accuracy for the regions close to the equilibrium geometry of a given molecular system, the same techniques fail for stretched bond distances. 
Similarly, the applicability of multi-reference CC or CI methods 
\cite{doi:10.1080/00268977500103351,kutzelnigg1982quantum,stolarczyk1985coupled,jeziorski1989valence,rittby1991multireference,kaldor1991fock,meissner1998fock,jezmonk,meissner1,pittnermasik,mahapatra1,mahapatra2,evangelista1,mrcclyakh}
is limited by the quality of model/active spaces that cannot be universally defined for all geometries involved in studies of chemical transformations usually related to bond-forming and bond-breaking processes. 
On the other hand, quantum algorithms are capable of alleviating issues with identifying electronic states of interest. An excellent illustration of this is the quantum phase estimation (QPE) method\cite{Kitaev:97,nielsen2002quantum,luis1996optimum, cleve1998quantum,berry2007efficient,childs2010relationship,seeley2012bravyi,wecker2015progress,haner2016high,poulin2017fast,berry2006}.
The probability of detecting a phase of a particular electronic state is proportional to the square of the overlap between the initial state used as a starting point for the unitary time evolution and exact wave function. 
Therefore, one can use the trial wave function as a form of the hypothesis to get the statistical echo of relevant exact states (and corresponding phases) that have a  non-zero overlap with the trial wave function.
Recently, we have tested these algorithms on the example of valence excited states and high-energy states corresponding to excitations of electrons occupying inner atomic shells.
When fault-tolerant quantum platforms are available, the QPE algorithms should be advantageous to the existing approximations for simulating core-level states on conventional computer platforms, especially in the context of identifying complex excited states dominated by multi-electron excitations.
In studies of excited-state processes, the QPE  approach has a clear advantage over the other algorithms, such as the Variational Quantum Eigensolver (VQE),
\cite{peruzzo2014variational,mcclean2016theory,romero2018strategies,PhysRevA.95.020501,Kandala2017,kandala2018extending,PhysRevX.8.011021,huggins2020non,cao2019quantum,anand2021quantum}
designed to utilize noisy intermediate-scale quantum (NISQ) devices, because detailed knowledge of the sought-after state is not required before the simulation.

The main focus of this paper is to highlight a recently developed QPE simulator that allows for scalable simulations of previously inaccessible atomic and molecular systems. We will discuss the implementation details,  which allow us to execute sizable QPE simulations for systems described by 10-20 molecular orbitals using modest computational resources provided by broadly accessible commodity computer systems. We will also illustrate the importance of this expanded QPE capability by studying the behavior of the ground and core-level energies as functions of the active-space size. As a benchmark system, we chose to investigate the ground and core-level excited states of H$_2$O. 

The paper is organized as follows: in Section 1, we provide an overview of the QPE algorithm. In Sections 2 and 3, we provide implementation details and a short description of the interface between the QPE algorithm and the NWChem \cite{valiev10_1477,apra2020nwchem} suite of computational chemistry software, which provides one- and two-electron integrals files and the second quantized characterization of the trial wave function, respectively. In Section 4, we discuss the results of the quantum algorithm simulations of H$_2$O, with special attention paid to the simulations of core-level excitations in the water molecule employing large active spaces. We demonstrate that using fifteen-orbital active space leads to a three-fold reduction of error in calculated core-level excitation energy discussed in Ref.~\citen{bauman2020toward}.
This result provides a strong argument for developing QPE algorithms for various types of x-ray spectroscopies and, in a more general perspective, identifying electronic states in arbitrary energy regimes.

\section{Methodology}
The main goal of quantum chemistry is to find high-quality approximate form of  solution to the electronic Schr\"odinger equation 
\begin{equation}
H |\Psi_K\rangle = E_K |\Psi_K\rangle \;,
\label{se1}
\end{equation}
where the electronic Hamiltonian $H$ depends parametrically on the coordinates of nuclei that form a fixed nuclear frame and $E_K$ and $|\Psi_K\rangle$ correspond to energy and wave function corresponding to $K$-th electronic state. 
The simulations of excited states usually involve a specific form of the excited-state wave function ansatz. For example, in the ubiquitous equation-of-motion coupled-cluster formalism (EOMCC),
\cite{bartlett89_57,bartlett93_414,stanton93_5178} the $K$-state is parametrized using the following Ansatz
\begin{equation}
|\Psi_K\rangle = R_K e^T |\Phi\rangle \;,
\label{eomcc}
\end{equation}
where $R_K$ and $T$ are many-body cluster and excitation operators and $|\Phi\rangle$ is the so-called reference function. To obtain accurate estimates of excitation energies for complex excited states, one needs to include high-rank collective excitations, which in many cases leads to insurmountable computational bottlenecks. Quantum computing and QPE formalism, in particular, have the potential to bypass these problems.

The language of second-quantization is not only a convenient language to express electronic Hamiltonian
\begin{equation}
     H = \sum_{p, q} h_{pq} a^{\dagger}_p a_q + \frac{1}{2}\sum_{p, q, r, s} h_{pqrs} a^{\dagger}_p a^{\dagger}_q a_r a_s
    \label{ham}
\end{equation}
and assure the anti-symmetry of the corresponding wave functions, but also gives rise to an efficient diagrammatic techniques for book-keeping the electron correlation effects of various order. The multi-dimensional tensors $h_{pq}$ and $h_{pqrs}$ are referred to as the one- and two-electron integrals.
In Eq.~(\ref{ham}), the creation/annihilation operators 
$a_p^{\dagger}$/$a_p$ for electrons in $p$-th single particle state (spin orbital) satisfy the standard Fermionic anti-commutation relations
\begin{eqnarray}
  &&[a_p,a_q]_+ = [a_p^{\dagger},a_q^{\dagger}]=0 \;,\label{comm1} \\
  &&[a_p,a_q^{\dagger}]_+ =\delta_{pq} \;, \label{comm2}
\end{eqnarray}
where $[A,B]_+=AB+BA$. Various spin orbitals bases can be used to express second quantized operators. In this paper, we will focus on the restricted Hartree-Fock (RHF) in calculations for ground and core-level states, and second-order MBPT natural orbitals (nMP2) in calculations for ground state energies. Assuming that we are dealing with system with $k_{\alpha}$ and $k_{\beta}$ $\alpha$ and $\beta$ electrons occupying $n$ orbitals, which corresponds to the wave function with defined $S_z$ quantum number, then the number of determinants in the FCI problem, $N_{\rm det}^{\rm (FCI)}$, corresponding to (\ref{se1}) is equal to \begin{equation}
    N_{\rm det}^{\rm (FCI)} = {n \choose k_{\alpha}}
    {n \choose k_{\beta}} \;.
    \label{dim1}
\end{equation}
This dimensionality can be further reduced by incorporating spin and spatial symmetry of the system of interest. However, for realistic systems, this leads to prohibitive costs for simulation on classical computers.

\subsection{Review of Hamiltonian Simulation via Quantum Phase Estimation}\label{qpe}
As broadly discussed in the literature, quantum algorithms that simulate the quantum evolution of a system described by Hamiltonian $H$ have asymptotically superior performance over classical {\it ab initio} techniques, enabling polynomial time and space complexity simulations. One technique, phase estimation-based Hamiltonian simulation, requires only a linear number of qubits relative to the number of orbitals \cite{Whitfield_2011}. We focus on phase estimation for its simplicity and utility.

In this section,
we first describe the phase estimation-based Hamiltonian simulation, including how Hamiltonians can be efficiently mapped to quantum architectures via the Jordan--Wigner transformation. Then, we discuss the role of the Lie--Trotter--Suzuki product formulas and error scaling. Finally, we describe how phase estimation can be used to extract the energy eigenvalue.

\subsubsection{Jordan--Wigner Transformation}
%
There are several transformations to map electronic Hamiltonians onto a register of qubits, including the Jordan--Wigner and Bravyi--Kitaev transformations \cite{Whitfield_2011, seeley2012bravyi}. Among them, the simplest (and historically oldest) is the Jordan--Wigner (JW) transformation that enables the representation of fermionic operators in terms of quantum gates:
\begin{align}
    a^{\dagger}_l = \bigotimes_{k = 1}^{l - 1} Z_k \frac{X_l - iY_l}{2} \qquad a_l = \bigotimes_{k = 1}^{l - 1} Z_k \frac{X_l + iY_l}{2} \;,
\end{align}
where $X, Y, Z$ are Pauli operators. 
Although the JW transformation is a non-local mapping in terms of fermionic degrees of freedom, it is sufficiently efficient in medium-scale applications to demonstrate the basic features of discussed quantum algorithms. 
The JW mapping transforms the second quantized Hamiltonian (\ref{ham}) into
a linear combination of unitaries (or Pauli strings) $P_k$
\begin{equation}
    H_{JW}=\sum_{k = 1}^m H_k = \sum_{k = 1}^m h_k P_k \;,
    \label{sou}
\end{equation}
where $h_k$ is a scalar factor. In our simulated systems, we assume that $h_k \in \mathbb{R}$, though there exist systems where $h_k \in \mathbb{C}$. 

\subsubsection{Trotterization}
Because $H_{JW}$ is Hermitian, $e^{-iH_{JW}t}$ is unitary and thus implementable on a quantum computer \cite{Whitfield_2011}. This operator can then be decomposed into a series of implementable unitaries on a quantum device via the Lie--Trotter--Suzuki matrix product formula \cite{hatano2005finding}. In particular, when $t \ll 1$, the first order non-symmetric Trotter formula can be applied as follows:
\begin{equation}
    \norm{e^{-iH_{JW}t} - \prod_{k = 1}^m e^{-iH_kt}} \in \mathcal{O}(m^2 t^2) \;,
\end{equation}
where $e^{-iH_k t}$ is implementable (the precise circuits required are further described in Ref.~\citen{Whitfield_2011}). Thus, the exponentiated second-quantized Hamiltonian can be simulated on a quantum computer, with quadratic error scaling. Furthermore, this technique can be generalized when we no longer can guarantee $t \ll 1$ via Trotterization. By selecting some number of Trotter steps $r$, we approximate the timesliced evolution operator $e^{-iH_{k}t/r}$ with $e^{i H_k t/r}$ terms, which has error scaling in:
\begin{align}
    \norm{e^{-iH_{JW}t} - \left( \prod_{k = 1}^m e^{-iH_k (t /r)} \right)^r} \in \mathcal{O}\left( \frac{m^2 t^2}{r} \right) \;.
\end{align}
Thus, by selecting $r$ appropriately with respect to the desired $t$ and the number of Hamiltonian terms $n$, an acceptable error bound can be derived. Furthermore, this was simply the first order, non-symmetric Trotter formula; higher-order formulas exist and have preferable asymptotic scaling at the cost of quantum operations \cite{berry2006}.

\subsubsection{Quantum Phase Estimation and Scaling}
Now that we can implement an approximation of $e^{-iHt}$ with constrained error, we must extract the energy eigenvalue from the operator. Recall that for any eigenvector $\ket{\psi}$ of the electronic Hamiltonian we have:
\begin{align}\label{eq:energy-phase}
    e^{-iHt} \ket{\psi} = e^{-iE_\psi t} \ket{\psi}
\end{align} where $E_\psi$ is the associated energy level. Thus, the core challenge is to extract the phase $E_\psi$ from the coefficient. In fact, this is precisely the objective of phase-estimation (PE) techniques \cite{svore2013faster}. PE operates such that, given a unitary $U$, eigenvector $\ket{\psi}$, and the number of precision bits $p$, we can extract the phase $\phi$:
\begin{align}\label{eq:qpe}
    U \ket{\psi} = e^{2 \pi i \phi} \ket{\psi} \;.
\end{align}
using an additional $p$ qubits for precision. 
Recognize that Eq.~\ref{eq:energy-phase} is precisely the form of Eq.~\ref{eq:qpe} when $2 \pi \phi = - E_\psi t$. Thus, we can apply PE to the Hamiltonian operator $e^{-iHt}$, and compute the energy as follows:
\begin{align}
    E_\psi = \frac{2 \pi}{t}  \phi \;.
\end{align}
When the time evolution operator is Trotterized, it typically is applied once as opposed to $r$ times. This still allows us to extract the energy:
\begin{align}
    e^{-iHt /r} \ket{\psi} = e^{- i E_\psi t / r} \ket{\psi} \implies E_\psi = \frac{-2 \pi r}{t} \phi  \;.
\end{align}
We typically take $t = 1$, so that $E_\psi = - (2 \pi r) \phi $. Thus, given a phase estimate from PE, we can produce an estimation of the energy eigenvalue.

In our implementation, we emulate a Fourier-based quantum phase estimation (QPE) \cite{abrams1999}. QPE requires $2^p - 1$ applications of $U$, $p + 2n$ qubits, where $2n$ corresponds to the number of spin orbitals, and applies the inverse quantum Fourier transform to produce $\phi$, which can be used to compute $E_\psi$.
\section{Algorithm}
The discussed implementation of  the QPE algorithm (\texttt{QPESIM}) has been developed using a C++ environment \cite{Kang_QPESIM_2022} and is optimized to take advantage of mid-range computer architectures. \texttt{QPESIM} is capable of performing QPE simulations for chemical systems and processes driven by ground and/or excited states in various energy regimes. In the following section, we provide details of \texttt{QPESIM}'s implementation.


\subsection{Psuedocode and Implementation}\label{subsec:psuedocode}
Our algorithm simulates QPE-based energy estimation on molecular Hamiltonians. \texttt{QPESIM} (Algorithm \ref{algo:simulate}) takes as input Hamiltonians produced by NWChem in the Broombridge format, a starting state to simulate, and two initial hyperparameters $p$, the number of bits of precision for phase estimation, and $r$, the number of Trotter steps. The algorithm then outputs a probability distribution over all phases from $0$ to $2^p - 1$:

\begin{algorithm}[H]
\DontPrintSemicolon 
\KwIn{A second quantized Hamiltonian $H$ with $n$ orbitals and $k_{\alpha}+k_{\beta} = 2 k_\alpha$ electrons stored as an array of classes of terms, $p$ number of ancilla precision bits, $r$ steps}
\KwOut{{\sc Probabilities}, a vector of length $2^p$ with the probability of each phase after QPE}
{\sc StateVector} $\gets$ complex, combinatorial ansatz vector of length $\binom{n}{k_{\alpha}}\binom{n}{k_{\alpha}} 2^p$\;
\For{$i = 0$; $i \leq 2^p - 1$; $i+= 1$} {
    {\sc ApplyUnitary}(i, {\sc StateVector})\;
}
{\sc Probabilities} $\gets$ InverseQuantumFourierTransform({\sc StateVector})\;
\Return{{\sc StateVector}}\;
\caption{{\texttt{SIMULATE}} emulates QPE-based Hamiltonian simulation}\label{algo:simulate}
\end{algorithm}


The key contribution is that $\texttt{StateVector}$ has an improved data representation that enables a superpolynomial reduction in space complexity. In Section~\ref{statevector-rep}, we discuss the data layout of \texttt{StateVector}. Then, in Section~\ref{statevector-update}, we describe the implementation of new loop update structures to execute \texttt{ApplyUnitary} on the novel data layout. 



\subsection{Representing Data in \texttt{StateVector}}\label{statevector-rep}
One of the limiting factors for typical quantum simulators is the na\"ive preparation of state vectors, often ignoring the symmetries of molecular systems. \texttt{QPESIM} improves upon this by being mindful of the chemical problem at hand and accounting for electronic symmetries when storing configuration amplitudes in \texttt{StateVector}. The benefit of the state preparation in \texttt{QPESIM} is fully appreciated when describing the algorithm using the occupation number representation.



\subsubsection{Occupation Number Representation}
For interacting fermionic systems, to describe the action of the creation/annihilation operators on the Slater determinants it is convenient to invoke the occupation number representation, where each Slater determinant is represented as a vector
\begin{equation}
|n_M \; n_{M-1}\;  \ldots\;  n_{i+1}\; n_i \;n_{i-1} \;\ldots \;n_1 \rangle
\label{onr}
\end{equation} 
where occupation numbers $n_i$ are equal to either 1 (electron occupies $i$-th spin orbital) or 0 (no electron is occupying $i$-th spin orbital). In (\ref{onr}), $M$ stands for the total number of spin orbitals used to describe quantum system and $M=2n$, where $n$ is the number of orbitals.  

The following formulas give the non-trivial action of creation/annihilation operators on the state vectors
\begin{widetext}
\begin{eqnarray}
a_i^{\dagger} |n_M \; n_{M-1} \; \ldots \;n_{i+1} \; 0 \; n_{i-1}\; \ldots n_1\; \rangle  &=& (-1)^{\sum_{k=1}^{i-1} n_k} 
|n_M \; n_{M-1} \; \ldots \;n_{i+1} \; 1 \; n_{i-1}\; \ldots n_1\; \rangle
\label{onr1} \;\;\;\;\;\; \\
a_i |n_M \; n_{M-1} \; \ldots \;n_{i+1} \; 1 \; n_{i-1}\; \ldots n_1\; \rangle  &=& (-1)^{\sum_{k=1}^{i-1} n_k} 
|n_M \; n_{M-1} \; \ldots \;n_{i+1} \; 0 \; n_{i-1}\; \ldots n_1\; \rangle .
\label{onr2}
\end{eqnarray} 
\end{widetext}
It should be stressed that the language of second quantization automatically assures proper anti-symmetry of the wave function expansions. 

Since the discussed implementation of the QPE  algorithm utilizes the list of the configurations in the occupation number representation, it is easy to incorporate various symmetries, including spin and spatial symmetries of the quantum system of interest. Another essential element of flexibility is the possibility of restricting the configurational space for the QPE simulations. For example, one can envision the process of selecting configurations that correspond to the Hartree-Fock determinant and all singly, doubly, triply, and quadruply excited Slater determinants with respect to the HF Slater determinant. This allows one to form the analogs of the truncated or even adaptive (multi-reference and/or state-selective) configuration-interaction expansions. In the example above, the selection of amplitudes would provide  CISDTQ 
(configuration interaction with singles, doubles, triples, and quadruples)
energies in the QPE simulations. 

\subsubsection{\texttt{QPESIM} approach to state vectors} 

Existing quantum simulators store all potential qubit configurations: assuming $n$ is the number of orbitals, simulators must store $2^M = 2^{2n}$ amplitudes. This is because each of the $2n$ spin orbitals are allocated a qubit in QPE, so each configuration from $00...0$ to $11...1$ must be stored. However, when accounting for symmetries and conservation of electrons, there are only at most $N_{\rm det}^{\rm (FCI)}$ valid configurations. Thus, using general-purpose quantum emulators for simulation algorithms often amounts to storing significantly more amplitudes than necessary to simulate the system. 

Our algorithm exploits configuration constraints by creating a vector \texttt{StateVector} which only stores valid configurations. The data structure itself can be represented as a 3D array defined by three indices: $l$, $c_1$, and $c_2$. In a given layer $(l, c_1, c_2)$, $l$ expresses the number of times the unitary $\prod_{j = 1}^n e^{-i H_j t / r}$ has been applied and $c_1$ and $c_2$ serve as indices for $\binom{n}{k_{\alpha}}$ alpha and $\binom{n}{k_{\beta}}$ beta
spin-orbital configurations in the amplitude vector, respectively. In our simulated system of H$_2$O with 15 orbitals, this exhibits about a 100x reduction in memory, and the savings would only be greater for larger systems. In general, our approximation has a memory reduction of at least $2^{M} / N_{\rm det}^{\rm (FCI)}$ compared to traditional quantum simulation approaches.

\subsection{Updating \texttt{StateVector}}\label{statevector-update}

While the \texttt{StateVector} format requires asymptotically less memory, it also requires a new update scheme. Matrix-based emulators simply use existing matrix product libraries to compute a series of matrix-vector products; in \texttt{QPESIM}, we instead iterate over the Hamiltonian terms and directly apply updates to the respective \texttt{StateVector} indices.


To achieve the update, we analyze the impacts of each type of Hamiltonian term on \texttt{StateVector}. Out of the five Hamiltonian term types, $pp$, $pq$, $pqqp$, $pqqr$, and $pqrs$, two classes can be created:
\begin{enumerate}
    \item \textbf{Configuration preserving} ($pp$, $pqqp$): these terms only modify the coefficient of configurations. Thus, the $pp$/$pqqp$ terms simply apply a multiplicative scalar coefficient
    \item \textbf{Configuration modifying} ($pq$, $pqqr$, $pqrs$): these terms represent interactions between two electronic configurations. Thus, these terms apply a 2x2 matrix transformation to an electron configuration and its conjugate.
\end{enumerate}


\subsubsection{Configuration Preserving Terms}
We first provide the effect of the configuration preserving operators. Suppose we have operators $a_p^\dagger$, $a_p$ operating on the spin orbital $p$ and an arbitrary state 
\begin{equation}
|\ldots \; n_p \; \ldots \rangle
    \label{pstate}
\end{equation}
where $p$-th spin orbital is occupied (i.e. $n_p=1$) then using definitions (\ref{onr1}) and (\ref{onr2}) the action of the $pp$ transformations ($a_p^{\dagger} a_p$) and their exponentials on the above vector is given by the formula
\begin{widetext}
\begin{equation}
    e^{-i (h_{pp} (a^{\dagger}_p a_p)) t} |\ldots \; n_p \; \ldots \rangle = 
    e^{-i (h_{pp} n_p) t} |\ldots \; n_p \;\; \ldots \rangle \;\; (n_p=1) \;.
    \label{ppex}
\end{equation}
\end{widetext}

Similarly, if in the state 
\begin{equation}
    |\ldots \; n_q\; \ldots \; n_p \; \ldots\rangle \;,
\end{equation}
spin orbitals $p$ and $q$ are occupied (i.e., $n_p=n_q=1$) then the action of the  $pqqp$ transformation on this vector can be written as:
\begin{widetext}
\begin{align}
    e^{-i (h_{pqqp} (a^{\dagger}_p a^{\dagger}_q a_q a_p)) t} 
    |\ldots \; n_q\; \ldots \; n_p \; \ldots\rangle
    &= e^{-i (h_{pqqp} n_p n_q ) t} |\ldots \; n_q\; \ldots \; n_p \; \ldots\rangle \;\; (n_p=n_q=1)
    \label{pqex}
\end{align}
\end{widetext}
This is a consequence of the fact that for $p\ne q$ the product $a_p^{\dagger} a_q^{\dagger} a_q a_p$
is equal to the product of particle number operators $\hat{n}_p \hat{n}_q$ where 
$\hat{n}_p=a_p^{\dagger}a_p$ and $\hat{n}_q=a_q^{\dagger} a_q$.


On all other states, these transformations simply apply identity. Thus, for the $pp$/$pqqp$ terms, \texttt{QPESIM} simply needs to find affected indices and multiply the associated coefficients by $e^{-ih_{pp} t /r}, e^{-ih_{pqqp} t/r}$, respectively. In code, this is implemented as:
\begin{widetext}
\begin{lstlisting}[language=C++]
StateVector[psi_pp_index] *= std::exp(-i * hpp * t)
StateVector[psi_pqqp_index] *= std::exp(-i * hpqqp * t)
\end{lstlisting}
\end{widetext}

\subsubsection{Configuration Modifying Terms}

The second class of terms, $pq$, $pqqr$, and $pqrs$, require
more complex updates. In contrast to the configuration preserving terms, the $pq$, $pqqr$, and $pqrs$ terms can produce singly or doubly excited configurations with respect to the configuration these terms are acting on. 

By using the Jordan--Wigner transformation, we can derive an explicit form for the 2x2 linear transformation given the coefficients of the two configurations. In particular, it's necessary to introduce $\eta$, the parity of the bits between the $p, q, r, s$ orbitals. For example, suppose $h$ is the Hamiltonian term coefficient of an $a_p^{\dagger} a_q^{\dagger} a_s a_r$ string and $\alpha_{\ket{01}}$ and $\alpha_{\ket{10}}$ are the coefficients of the conjugate configurations for that specific term. Then, the transformation can be described via a linear, symmetric matrix:
\begin{align}
    \begin{bmatrix}
    \cos (h t) & (-1)^{\eta + 1} i \sin (h t) \\
    (-1)^{\eta + 1} i \sin (h t) & \cos (h t)
    \end{bmatrix}
    \begin{bmatrix}
    \alpha_{\ket{01}} \\
    \alpha_{\ket{10}}
    \end{bmatrix}
\end{align}


This simplified translation also reduces the complexity of the code, as all terms can be reduced to accessing at most two elements of the state vector. Thus, the following code is sufficient:
\begin{widetext}
\begin{lstlisting}[language=C++]
void updateStateVector(std::array<std::complex<double>, 4> coeffs, 
        unsigned leftIdx, unsigned rightIdx)
{
    stud::complex<double> left, right;
    left = StateVector[leftIdx];
    right = StateVector[rightIdx];
    
    StateVector[leftIdx] = left * coeffs[0] + right * coeffs[1];
    StateVector[rightIdx] = left * coeffs[1] + right * coeffs[0];
}

\end{lstlisting}
\end{widetext}

\subsubsection{\texttt{ApplyUnitary}  Pseudocode}
In order to identify the indices to modify, we employ the following loop hierarchy in the \texttt{ApplyUnitary} method (Algorithm \ref{algo:apply_unitary}):
\begin{enumerate}
    \item Class of Hamiltonian term (for each type $pp$, $pqqp$, $pq$, $pqqr$, $pqrs$, there is a list of Hamiltonian terms $H_{pp}, H_{pqqp}, H_{pq}, H_{pqqr}, H_{pqrs}$)
    \item Term within the class
    \item Affected amplitudes/state vector indices
\end{enumerate}

This produces the following update algorithm:
\begin{algorithm}
\DontPrintSemicolon 
\KwIn{Sets of terms broken by class, $H_{pp}, H_{pqqp}, H_{pq}, H_{pqqr}, H_{pqrs}$ and a starting ansatz {\sc StateVector}}
\KwOut{$e^{-iHt}$ is applied once to {\sc StateVector}}
\For{{\sc Class} in $ \{H_{pp}, H_{pqqp}, H_{pq}, H_{pqqr}, H_{pqrs}\}$} {
    \For{{\sc Term} in {\sc Class}}{
        \For{{\sc Configuration} affected by {\sc Term}} {
            {\sc IndicesToUpdate} $\gets$ ConfigurationToIndex({\sc Configuration})\;
            UpdateStateVector({\sc IndicesToUpdate}, {\sc Term}.{coefficients})\;
        }
    }
}
\caption{{\sc ApplyUnitary} applies the $e^{-iHt}$ unitary once}
\label{algo:apply_unitary}
\end{algorithm}


Note that we update the state vector in a particular ordering, as specific Hamiltonian orderings affect the accuracy of the Trotter--Suzuki formula\cite{hastings2014improving}.

\section{Results \& Discussion}
We chose the H$_{2}$O as an exemplary system that we could examine both ground and high-energy excited 
states with while going beyond the capabilities of alternative QPE solvers. The geometry of the system is given by $R_{\rm OH} = 0.95778$ \r{A} and $\theta_{HOH}=104.47984^{\circ}$. In our earlier study \cite{bauman2020toward},
we demonstrated the utility of the QPE algorithm in targeting core-level excited states. However, we were
restricted to stinted active spaces that lead to understandably imperfect results when compared to experimental results. In this study, we employed an active space consisting of the 15 lowest-lying RHF orbitals and MP2 natural orbitals of the cc-pVDZ basis set (24 orbitals in the full basis).

\begin{figure}
    \centering
    \includegraphics[scale=0.5]{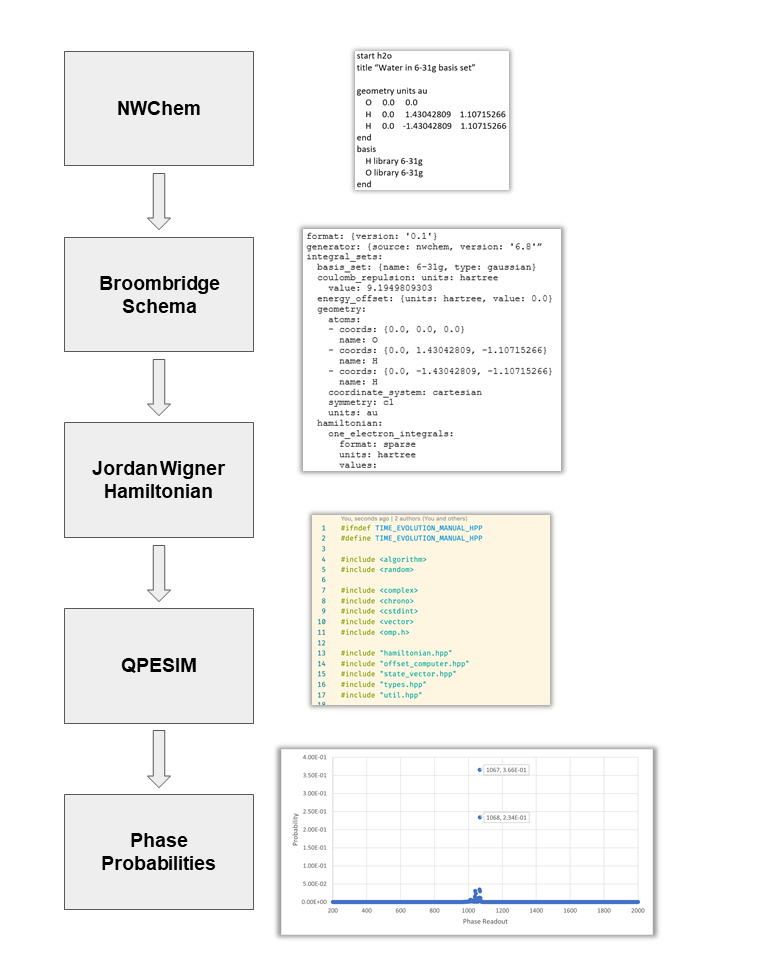}
    \caption{Flowchart describing the \texttt{QPESIM} simulation process. Systems are represented using NWChem and the corresponding Hamiltonian is extracted using the Broombridge Schema. Next, this Hamiltonian can be converted into an implementable Hamiltonian via the Jordan--Wigner transformation. QPESIM ingests this Hamiltonian to produce a series of phase probabilities, i.e. the probability of obtaining a certain phase from QPE. These phase probabilities can be used to produce an energy level estimate for the associated ansatz.}
    \label{fig:flowchart}
\end{figure}

Our primary contribution, a new algorithm for quantum phase estimation-based simulation, requires significantly less memory than general purpose quantum simulators and integrates with NWChem and the Broombridge Hamiltonian format. Our end-to-end pipeline, as described in Figure \ref{fig:flowchart}, first ingests Broombridge Hamiltonians from NWChem, converts it to a Jordan--Wigner Hamiltonian, and runs the simulation. The final output is a distribution of phases corresponding to the predicted energy level of the system.

\begin{figure}
    \centering
    \includegraphics[scale=0.3]{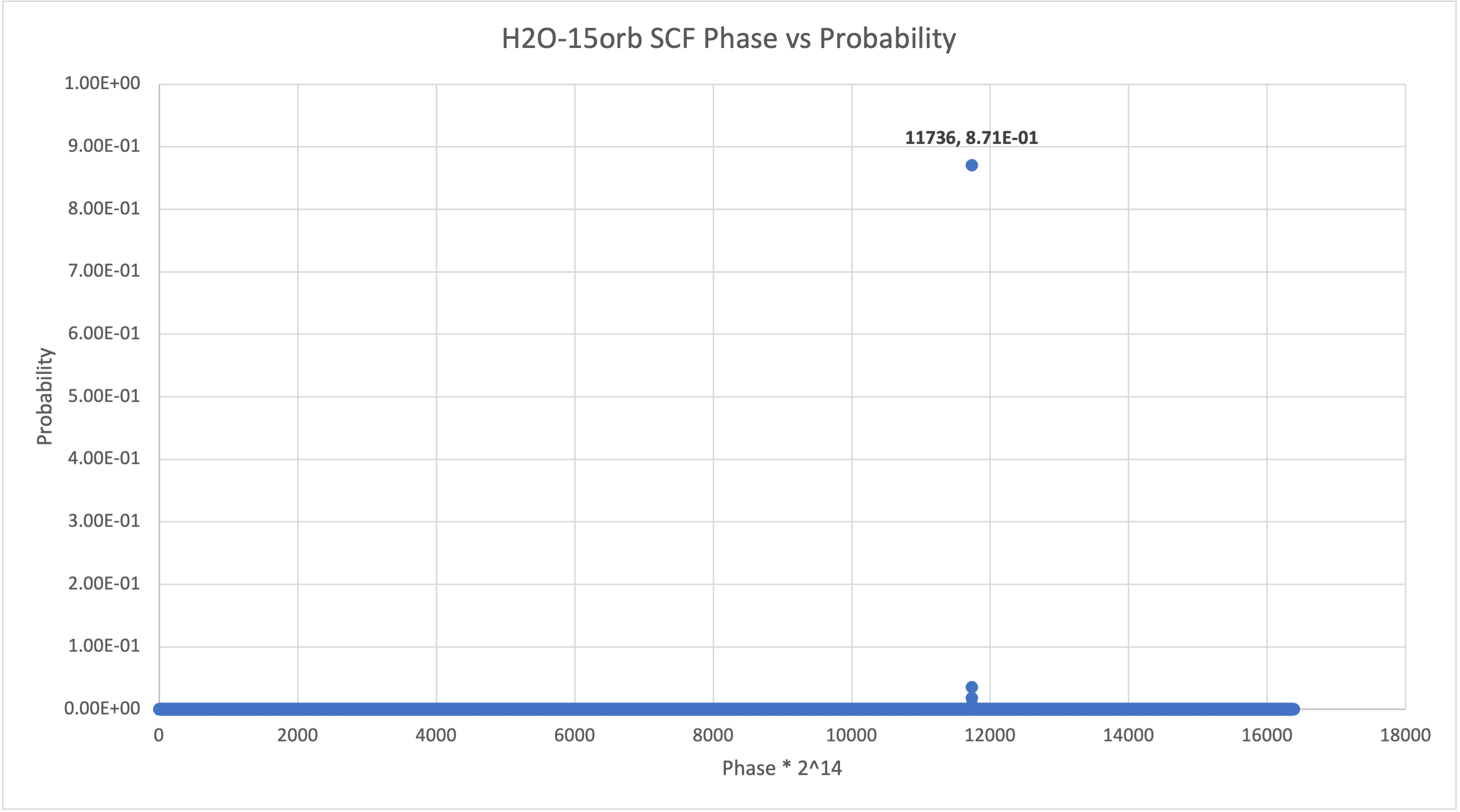}
    \caption{The phase/probability distribution for the ground state of H$_{2}$O, obtained using the 15 lowest-energy RHF orbitals. There is a primary peak at 11736 with probability 87.1\%.}
    \label{fig:scf-ground}
\end{figure}

\begin{figure}
    \centering
    \includegraphics[scale=0.3]{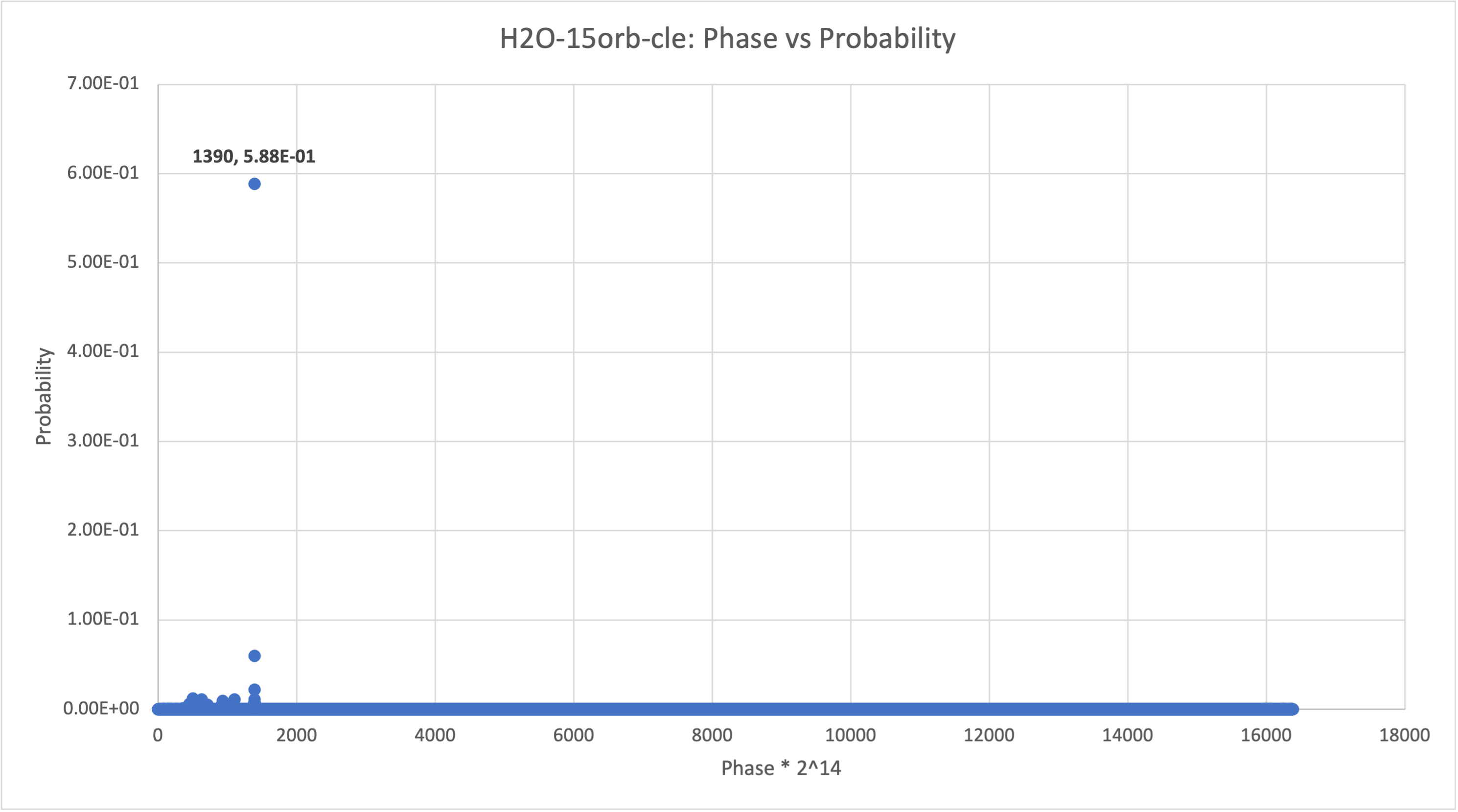}
    \caption{The phase/probability distribution for the lowest-energy core-level excited state of H$_{2}$O ($1a_1\rightarrow 4a_1$ transition), obtained using the 15 lowest-energy RHF orbitals. We observe a dominant phase of 1390 with probability 58.8\% which was taken as the lone peak in our computation of the energy level.}
    \label{fig:scf-excited}
\end{figure}

\begin{figure}
    \centering
    \includegraphics[scale=0.3]{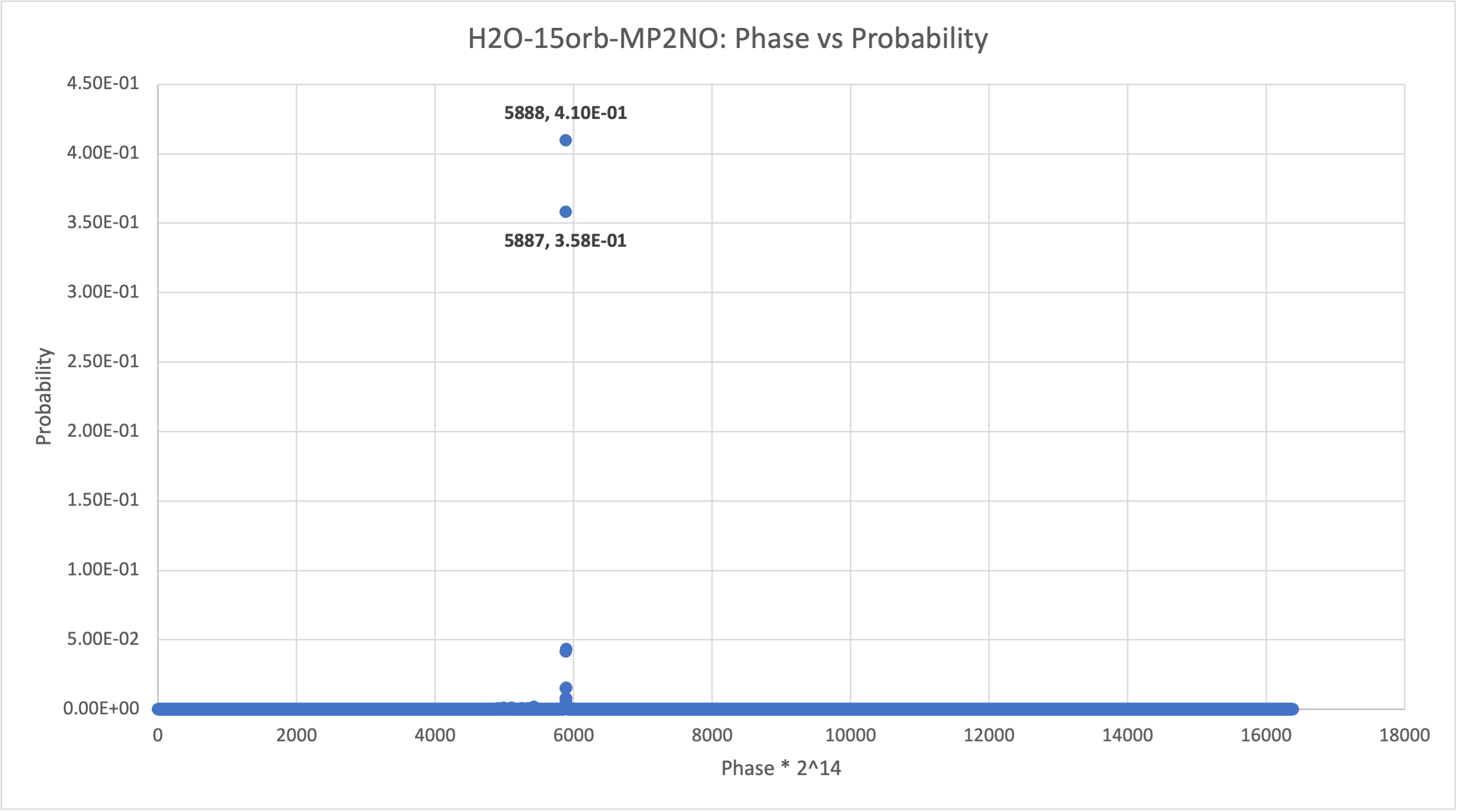}
    \caption{The phase/probability distribution for the ground state of H$_{2}$O, obtained using the 15 lowest-energy MP2 natural orbitals. We observe that the phase is split between two readouts: 5888 with probability 41.0\% and 5887 with probability 35.8\%. Table \ref{tab:energies} contains the corresponding energies for these two phases, as well as the weighted average of these two readouts.}
    \label{fig:mp2-ground}
\end{figure}

When we view the phase distributions on our sample systems (see Figures \ref{fig:scf-ground}-\ref{fig:mp2-ground}), the phase results reflect how well our ansatz approximates the target-state eigenvector, as seen in the probability distribution of the phases. Essentially, while the Trotter formula does slightly modify the true eigenvectors, we would expect an ansatz with high overlap with the original target-state configuration to also have a phase readout with a high probability of measuring the correct energy phase. In particular, by taking the phase(s) with the highest probability, we can produce an energy level estimate of the system.

As seen in Table \ref{tab:energies}, the 15 orbital ground-state QPE simulation is 
relatively close to the analogous classical exact (FCI) solution and is lower 
than the previous 9 orbital simulations by $\sim$0.1 Hartree. The error relative to 
the full cc-pVDZ CCSDTQ (CC with singles, doubles, triples, and quadruples)
\cite{ccsdtq_nevin,Kucharski1991,doi:10.1063/1.467143}
benchmark is reduced by more than a factor of three for 
the 15 orbital simulation compared to the previous smaller active space 
simulation.  For the excited state, the larger active space lowers the total energy by 0.2--0.3 Hartree. As a result of the improved 
energies for the ground and excited state energies, the error of the 
excitation energy for the larger active space is reduced by more than a 
factor of two when compared to experiment and is only a few eVs higher than the 
benchmark EOMCCSD (Equation-of-motion CC with singles and doubles) 
\cite{bartlett89_57,bartlett93_414,stanton93_5178}
and EOMCCSDT
(EOMCC with singles, doubles, and triples) \cite{kkppeom}
excitation energies obtained with the full cc-pVDZ 
orbital space. We would also like to note that while previous simulations and EOMCC benchmarks used a slightly different geometry, the differences in energies due to the changes in geometry are insignificant compared to the change in energies due to the larger active space.

While larger active spaces can significantly improve results and expand the capabilities of quantum simulations, there is much value in techniques for compacting electronic correlation and reducing the dimensions of a problem through various techniques. A simple demonstration of this is seen when we replace the RHF orbitals in the active space with the 15 largest occupation MP2 natural orbitals for the ground state. Not only is the energy reduced by approximately 60 milliHartree compared to the analogous RHF case, but the improved energy is only 23 milliHartree from the benchmark CCSDTQ energy for the full orbital space. This error would further improve with natural orbitals from higher-order methods and approaches such as the downfolding techniques that we introduced in earlier studies.

\begin{table}
    \centering
    \begin{tabular}{l c c }
        \hline \hline \\
        Method & Orbitals & Energy \\[0.2cm]
        \hline \\
        \multicolumn{3}{c}{\underline{Ground State}} \\[0.1cm]
        QPE$^{a}$ & RHF (9 Orbitals)  & -76.0591   \\[0.1cm]
        QPE(p=14, r=5) & RHF (15 Orbitals)  & -76.1612  \\[0.1cm]
        QPE(p=14, r=10)$^b$ & nMP2 (15 Orbitals)  & -76.2225\\[0.1cm]
        & & -76.2187 \\[0.1cm]
        & & (-76.2207)  \\[0.2cm]
        FCI & RHF (15 Orbitals)  & -76.1482   \\[0.1cm]
        CCSDTQ$^{c}$ & RHF (24 Orbitals)  &  -76.2438  \\[0.2cm]
        \multicolumn{3}{c}{\underline{Core-level Excited State}} \\[0.1cm]
        QPE$^{a}$ & RHF (9 Orbitals)  & -55.9517   \\[0.1cm]
        QPE(p=14, r=5) & RHF (15 Orbitals) & -56.3076
        \\[0.2cm]
        \multicolumn{3}{c}{\underline{Excitation Energies}} \\[0.1cm]
        QPE$^{a}$ & RHF (9 Orbitals)  &  547.15  \\[0.1cm]
        QPE(p=14, r=5) & RHF (15 Orbitals) &  540.24  \\[0.1cm]
        EOMCCSD$^{d}$ & RHF (24 Orbitals) & 538.40 \\[0.1cm]
        EOMCCSDT$^{e}$ & RHF (24 Orbitals) & 537.32 \\[0.1cm]
        Experiment$^{f}$ &  -- & 534.0 \\[0.2cm]
%
        \hline \hline
    \end{tabular}
    \caption{Total energies (Hartree) for the ground ($1^1A_1$) and excited 
    ($1a_1\rightarrow 4a_1$ transition)
    states of H$_{2}$O, along with their respective excitation energies (eV).}
    \label{tab:energies}
\footnotesize{$^{a}$Taken from Ref.~\citen{bauman2020toward}.}
\footnotesize{$^{b}$Two neighboring phases with probabilities greater than 0.1 were identified within the same simulation. Both energies are reported along with a weighted average, in parenthesis, based on their respective probabilities.}
\footnotesize{$^{c}$The CCSDTQ total energy for the full cc-pVDZ orbital space.} 
\footnotesize{$^{d}$The EOMCCSD cc-pVDZ excitation energy taken from Ref.~\citen{brabeccorelvl}.}
\footnotesize{$^{e}$The EOMCCSDT cc-pVDZ excitation energy taken from Ref.~\citen{brabeccorelvl}.}
\footnotesize{$^{f}$Experimental excitation energy taken from Ref.~\citen{H2Oexpcorelvl}}
\end{table}

\section{Conclusion}
In this paper, we demonstrated the versatility of the \texttt{QPESIM} code in applications to chemical problems. Our primary focus was on describing electronic states corresponding to various energy regimes. As a particular example, we chose the problem of calculating excitation energies corresponding to core-level excitation of the water molecule recently analyzed using the QPE algorithm. Our \texttt{QPESIM} implementation allowed for the utilization of significantly larger active spaces than previously possible. 
As a result, when using active space defined by 15 RHF orbitals, we could observe a significant two-fold reduction of computed core-level excitation energy error obtained in QPE simulations employing 9 RHF active orbitals when compared with experiment. The obtained QPE results for core-level excitation energy corresponding to the singly-excited core-level  state of $^1A_1$ symmetry are in good agreement with the EOMCCSD results obtained with all orbitals being correlated. 

In the \texttt{QPESIM} calculations for the ground-state energy of water molecules in equilibrium configurations, we also demonstrated the advantage of utilizing natural orbitals determined at the MP2 level. Using active space composed of 15 natural active orbitals, we achieved 
more than a three-fold reduction of error in correlation energy obtained in simulations involving 15 RHF active orbitals
compared to the CCSDTQ results obtained for all orbitals being correlated. 

There is an ongoing effort towards further optimization of the \texttt{QPESIM} code, which should enable QPE simulations of systems described by 20-25 active orbitals, which will enable us for the direct evaluation of the QPE performance  compared to the results obtained with high-order CC/EOMCC  methods  for a broad class of electronic states.

\section{Acknowledgement}
The main part of this  work was supported by  the Quantum Science Center (QSC), a National Quantum Information Science Research Center of the U.S. Department of Energy (DOE).
Part of this work was supported by the "Embedding QC into Many-body Frameworks for Strongly Correlated  Molecular and Materials Systems'' project, which is funded by the U.S. Department of Energy, Office of Science, Office of Basic Energy Sciences (BES), the Division of Chemical Sciences, Geosciences, and Biosciences. C.K. thanks Nathan Wiebe for discussions about the nature of QPE-based algorithms for simulation.

\clearpage



\providecommand{\latin}[1]{#1}
\makeatletter
\providecommand{\doi}
  {\begingroup\let\do\@makeother\dospecials
  \catcode`\{=1 \catcode`\}=2 \doi@aux}
\providecommand{\doi@aux}[1]{\endgroup\texttt{#1}}
\makeatother
\providecommand*\mcitethebibliography{\thebibliography}
\csname @ifundefined\endcsname{endmcitethebibliography}
  {\let\endmcitethebibliography\endthebibliography}{}

\end{document}